\documentclass[conference]{IEEEtran}
\IEEEoverridecommandlockouts
\usepackage{cite}
\usepackage{amsmath,amssymb,amsfonts}
\usepackage{graphicx}
\usepackage{textcomp}
\usepackage{xcolor}
\def\BibTeX{{\rm B\kern-.05em{\sc i\kern-.025em b}\kern-.08em
    T\kern-.1667em\lower.7ex\hbox{E}\kern-.125emX}}
\usepackage{mathtools}
\usepackage{amsthm}
\usepackage{stfloats}
\usepackage{epstopdf}
\usepackage[linesnumbered,ruled,vlined]{algorithm2e}

\usepackage{subcaption}
\usepackage{algcompatible}
\usepackage{ragged2e}
\usepackage{float}
\usepackage[normalem]{ulem}	
\usepackage{soul}

\begin{document}


\title{Affine Frequency Division Multiplexing: From Communication to Sensing}
\author{
    \IEEEauthorblockN{Ali Bemani\IEEEauthorrefmark{1}, Nassar Ksairi\IEEEauthorrefmark{1} and Marios Kountouris\IEEEauthorrefmark{2}\IEEEauthorrefmark{3}}
    
    \IEEEauthorblockA{\IEEEauthorrefmark{1}Mathematical and Algorithmic Sciences Lab, Huawei France R\&D, Paris, France\\
    \IEEEauthorrefmark{2}Communication Systems Department, EURECOM, Sophia Antipolis, France\\
    \IEEEauthorrefmark{3}Department of Computer Science and Artificial Intelligence, University of Granada, Spain
    \\ali.bemani,nassar.ksairi@huawei.com, mariosk@ugr.es}

}

\maketitle

\begin{abstract}
Affine Frequency Division Multiplexing (AFDM) has been proposed as an effective waveform for achieving the full diversity of doubly-dispersive (delay-Doppler) channels. While this property is closely related to range and velocity estimation in sensing, this article focuses on other AFDM features that are particularly relevant for addressing two challenges in integrated sensing and communication (ISAC) systems: (1) maintaining receiver complexity and energy consumption at acceptable levels while supporting the large bandwidths required for high delay/range resolution, and (2) mitigating interference in multiradar environments. In monostatic sensing, where direct transmitter-receiver leakage is a major impairment, we show that AFDM-based ISAC receivers can address the first challenge through their compatibility with low-complexity self-interference cancellation (SIC) schemes and reduced sampling rates via analog dechirping. In bistatic sensing, where such analog solutions may not be feasible, we demonstrate that AFDM supports sub-Nyquist sampling without requiring hardware modifications while preserving delay resolution. Finally, we show that the second challenge can be addressed by leveraging the resource-assignment flexibility of the discrete affine Fourier transform (DAFT) underlying the AFDM waveform.
\end{abstract}


\section{Introduction}
Integrated sensing and communication (ISAC) is a key technology for next-generation wireless networks, particularly relevant to emerging applications such as vehicular networks, industrial Internet of Things (IoT), and smart homes, where seamless integration of sensing and communication is essential. Achieving high-resolution sensing in ISAC requires large signal bandwidths, since the achievable range resolution is inversely proportional to the available bandwidth. For instance, achieving a range resolution better than 10~cm demands a signal bandwidth and analog-to-digital converter (ADC) sampling rate exceeding 1.5~GHz. Such requirements impose significant challenges in terms of complexity, cost, and power consumption \cite{power_saving}, as the ADC must operate at a sufficiently high rate to capture the full-band signal. The challenge becomes even more severe in MIMO radar systems, where multiple transmit–receive channels \cite{han2022high} drastically increase the ADC demands. Furthermore, in dense deployments or multiradar environments, co-channel interference between radars can severely degrade sensing performance unless effective interference mitigation techniques are applied. These considerations motivate the design of ISAC solutions that lower the sampling rate while preserving sensing accuracy, and that can also robustly mitigate multiradar interference.

In monostatic ISAC systems, where the transmitter and receiver are colocated, the challenges of high sampling rates and multiradar interference are compounded by another critical issue: self-interference (SI). The strong coupling of the transmitted signal directly into the receiver can be several orders of magnitude stronger than the echoes from distant targets, creating a severe dynamic range problem for the receiver’s ADC. This issue is particularly problematic when using practical, low-resolution ADCs, as the SI can saturate the front end and mask the much weaker target returns. To preserve sensing performance, effective self-interference cancellation (SIC) must be applied before the ADC stage, ensuring that SI power is sufficiently suppressed to maintain sensitivity to reflected signals.

In conventional radar systems, the frequency-modulated continuous-wave (FMCW) waveform enables a low-complexity analog-domain SIC method via \emph{chirp carrier mixing}, an operation that not only suppresses SI before the ADC but also naturally reduces the required sampling rate through analog dechirping \cite{jin2021fmcw}. However, conventional approaches to mitigating multiradar interference in FMCW, such as multi–chirp-rate assignment \cite{wang2024interference}, are not well suited to cellular ISAC systems. Managing multicell ISAC interference would require the availability of a large pool of chirp-rate values, which would negatively impact both SIC performance and ADC sampling-rate requirements\footnote{The bandwidth of the received signal after dechirping is proportional to the chirp rate.}.

{\bf In the first part} of this magazine article, we present a unified approach to address three major challenges in {\bf monostatic sensing}: high ADC sampling rates, multiradar interference, and SI. The core of our solution is the use of \emph{Affine Frequency Division Multiplexing} (AFDM), a waveform based on the discrete affine Fourier transform (DAFT). In our earlier work \cite{bemani2023affine_TWC, bemani2021afdm}, we showed that AFDM achieves full diversity in doubly-dispersive wireless channels, making it a strong candidate for ISAC \cite{rou2025chirp}. More importantly, AFDM enables the use of chirp carrier mixing, as in FMCW radars, for both SIC and sampling-rate reduction. Furthermore, AFDM’s flexibility in pilot design allows the use of \emph{linear combinations of chirps} instead of assigning distinct chirp rates to different radars, thereby mitigating multiradar interference while keeping the chirp rate fixed across all systems. This eliminates the need for higher sampling rates, overcoming a key limitation of traditional FMCW-based interference management. However, the use of linear combinations of chirps introduces a trade-off: it creates a \emph{blind region} in which the monostatic radar cannot operate.

While analog-domain solutions such as chirp carrier mixing can be effective for SIC in monostatic ISAC systems, they may not be suitable for all deployment scenarios. In particular, when sensing is performed on the user equipment (UE) side, as in bistatic ISAC configurations, introducing additional analog processing is highly undesirable. UEs are typically constrained by hardware complexity, power consumption, and backward compatibility, which makes purely digital sub-Nyquist sampling methods more attractive. Although sub-Nyquist radar techniques have been widely studied~\cite{Eldar_sub_Nyquist}, many rely on specialized analog components such as multichannel mixers~\cite{Eldar_sub_Nyquist}, signal differentiators~\cite{superresolution_subNyquist}, or analog delay modules and RF splitters~\cite{hardware_subNyquist}, all of which increase implementation complexity and reduce robustness. These analog-intensive approaches are difficult to integrate into existing UE hardware platforms. Therefore, a digital sub-Nyquist solution that requires no changes to the analog front end is essential for practical and scalable UE-side sensing.

To completely eliminate the need for special analog hardware prior to ADC, random sampling has been proposed \cite{Eldar_sub_Nyquist}. Random sampling solutions exploit channel sparsity by using random (as opposed to uniform) sub-Nyquist sampling to achieve lower coherence in the sensing matrix, thus improving compressed sensing performance. However, nonuniform sampling approaches are often infeasible in real-world applications due to hardware limitations \cite{hardware_subNyquist}. In addition, several uniform sub-Nyquist sampling techniques have been explored to address the challenges of high-bandwidth radar processing, albeit with limitations. For example, stepped-carrier OFDM sequentially samples the full bandwidth using lower-bandwidth OFDM blocks in successive time intervals at different carrier frequencies \cite{schindler2019integrated}, but this complicates synchronization, increases latency and receiver power consumption, and reduces the maximum unambiguous velocity. Subcarrier-aliasing (SA)-OFDM \cite{lang2022ofdm} mitigates aliasing through carefully designed active-subcarrier intervals, simplifying hardware at the cost of degraded range resolution and spectral efficiency due to the need to deactivate inactive subcarriers. Sub-Nyquist OFDM radar \cite{han2023sub} downsamples the received signal and reconstructs it via an unfolding process, eliminating hardware overhead and Doppler ambiguities. However, it introduces symbol mismatch noise, which requires complex iterative noise cancellation.

{\bf In the second part} of this article, we introduce a purely digital sub-Nyquist {\bf bistatic sensing} solution based on the AFDM waveform. Unlike existing methods that rely on sparse nonuniform sampling, subcarrier interleaving, or dedicated analog-domain processing such as mixing and integration, our approach employs uniform sub-Nyquist sampling and requires no additional hardware modifications. This makes it well-suited for practical deployment, particularly in systems where simplicity, scalability, and backward compatibility are essential. We demonstrate that the proposed scheme preserves the delay resolution of the full-rate receiver, while the maximum unambiguous detection range is reduced in proportion to the downsampling factor, a manageable trade-off in many ISAC applications.

\section{AFDM Waveform and Its Role in ISAC Systems}
AFDM is a recently proposed multicarrier waveform that offers inherent resilience to both delay and Doppler impairments. It is constructed using the DAFT, which generalizes the discrete Fourier transform (DFT) by introducing time–frequency chirping into its basis functions. Let the signal occupy a bandwidth $\mathrm{BW}$~Hz with sampling interval $T_{\rm s} = 1/\mathrm{BW}$. The $N$-sample AFDM time-domain signal $X[n]$ is obtained from its DAFT-domain symbols $\{x_m\}$ via the inverse DAFT (IDAFT) as
\begin{equation}
\label{eq:td_afdm}
X[n] = \sum_{m \in \mathcal{P}} x_m,\phi_{m}[n], \quad n = 0, 1, \dots, N-1,
\end{equation}
where $\mathcal{P}$ denotes the set of pilot or data symbol indices, and each AFDM basis function is a discrete chirp of the form $\phi_{m}[n] \triangleq e^{\imath 2\pi \left(c_1 n^2 + c_2 m^2 + \frac{m n}{N} \right)}$, with $c_1$ and $c_2$ being the time- and frequency-chirp-rate parameters, respectively. This chirp-based subcarrier structure spreads the signal energy across the time–frequency plane, providing robustness in doubly-dispersive channels—an essential requirement for joint communication and sensing in high-mobility environments.

Unlike OFDM, which is highly sensitive to time and frequency offsets, AFDM maintains orthogonality under strong dispersion, enabling reliable performance in high-mobility scenarios. Previous studies~\cite{bemani2021afdm, bemani2023affine_TWC} have shown that AFDM achieves full diversity gains and improved bit-error-rate performance compared to OFDM and OTFS, underscoring its potential for next-generation wireless systems . From an ISAC perspective, AFDM’s resemblance to linear-frequency-modulated (LFM) waveforms makes it naturally compatible with radar processing methods for range–Doppler estimation, while its DAFT-domain representation supports flexible pilot and frame designs. Together, these properties position AFDM as a compelling waveform for unified communication and sensing, capable of meeting the stringent requirements of future integrated networks.

\section{AFDM-based Monostatic Systems}

\begin{figure}
\centering
    \includegraphics[width= \linewidth]{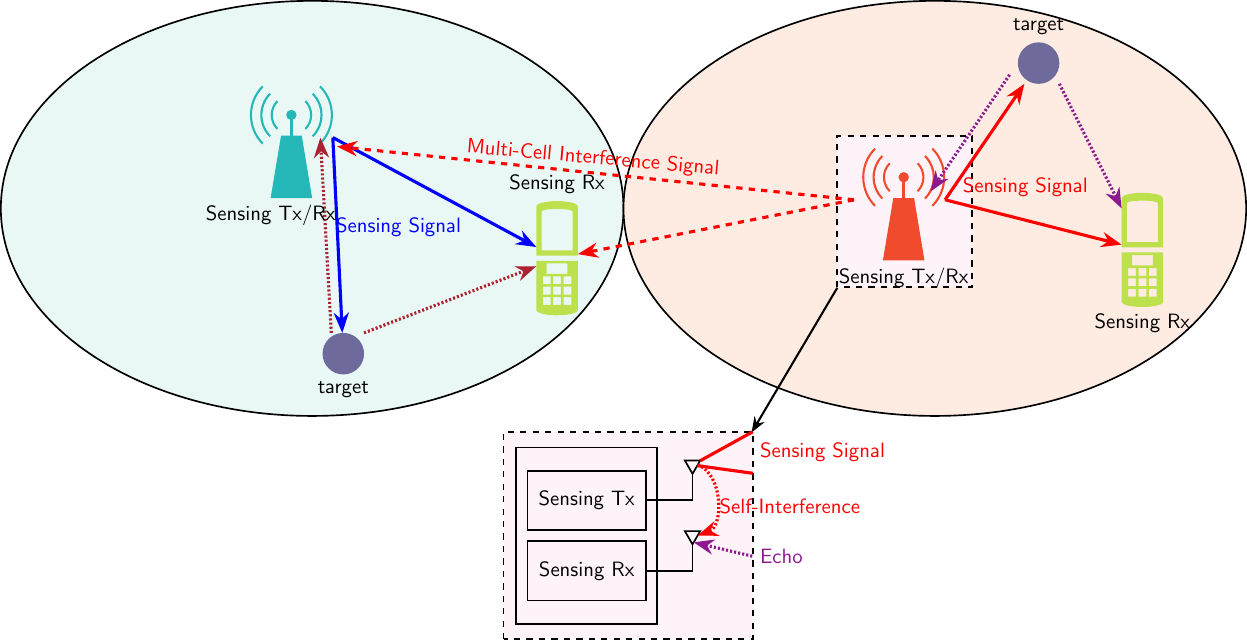}
\caption{System model for the sensing scenario} 
\label{fig:SysModel}
\end{figure}

\begin{figure}
\centering
\begin{subfigure}{.45\textwidth}
   \centering
    \includegraphics[width= \textwidth]{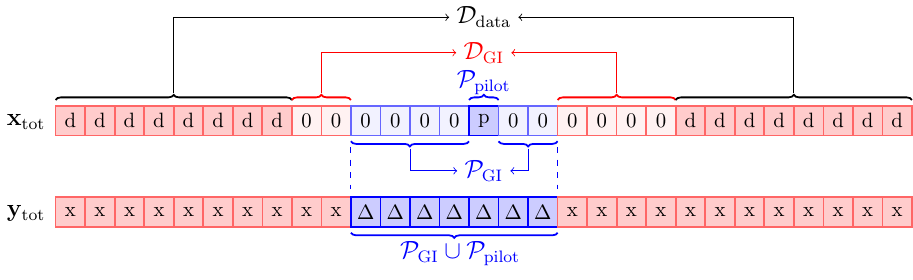}
    \caption{Monostatic sensing with one pilot multiplexed with data}
    \label{fig:FrameSignlePilotWithData}
\end{subfigure}
\hspace{0.1\textwidth}
\begin{subfigure}{0.45\textwidth}
\centering
    \includegraphics[width= \textwidth]{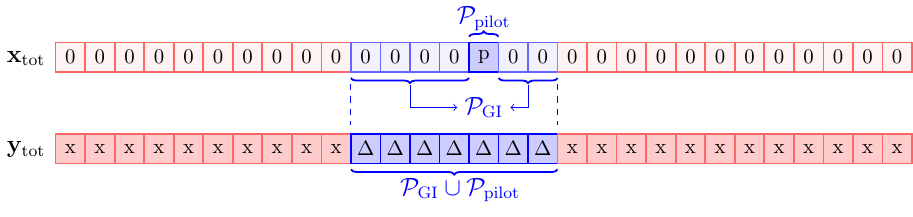}
    \caption{Bistatic sensing with one pilot}
    \label{fig:FrameSignlePilotWithoutData}
\end{subfigure}
\hspace{0.1\textwidth}
\begin{subfigure}{0.45\textwidth}
\centering
    \includegraphics[width= \textwidth]{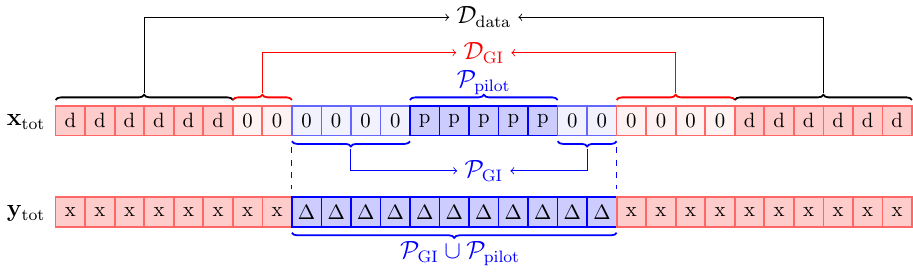}
    \caption{Monostatic sensing with linear combination of pilots multiplexed with data}
    \label{fig:FrameMultiplePilotWithData}
\end{subfigure}
\hspace{0.1\textwidth}
\begin{subfigure}{0.45\textwidth}
    \centering
    \includegraphics[width= \textwidth]{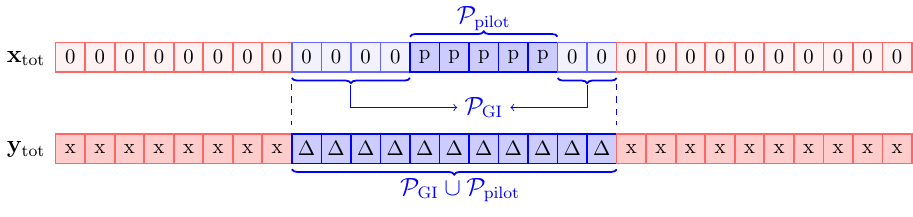}
    \caption{Bistatic sensing with linear combination of pilots}
    \label{fig:FrameMultiplePilotWithoutData}
\end{subfigure}
\caption{Transmitted and received AFDM frame} 
\label{fig:Frame}
\end{figure}

Fig.~\ref{fig:SysModel} illustrates an ISAC system that incorporates both monostatic and bistatic sensing configurations. In the monostatic setup, a colocated transmitter and receiver jointly perform communication and radar sensing at the base station. 
The monostatic transceiver employs AFDM. As illustrated in Fig.~\ref{fig:FrameSignlePilotWithData}, the AFDM signal can be structured to embed a pilot symbol within a DAFT-domain frame, accompanied by a carefully designed guard interval. This structure enables the receiver to resolve both delay and Doppler components of the wireless channel from a single symbol. The resulting delay–Doppler profile, extracted from the received echoes, directly maps to the range and velocity of nearby targets, making it naturally suited for sensing.

The remaining indices are used for communication data, maintaining orthogonality with the sensing operation. An additional benefit of the guard intervals is that they concentrate the transmit energy around the pilot, effectively boosting its power relative to the surrounding data. This power boost enhances the reliability of delay–Doppler estimation, particularly in low-SNR or highly-dispersive environments. Overall, this dual-purpose structure enables simultaneous communication and radar sensing within the same time–frequency resource, without the need for separate training sequences or mode switching.

\begin{figure}
    \centering
    \includegraphics[width=\linewidth]{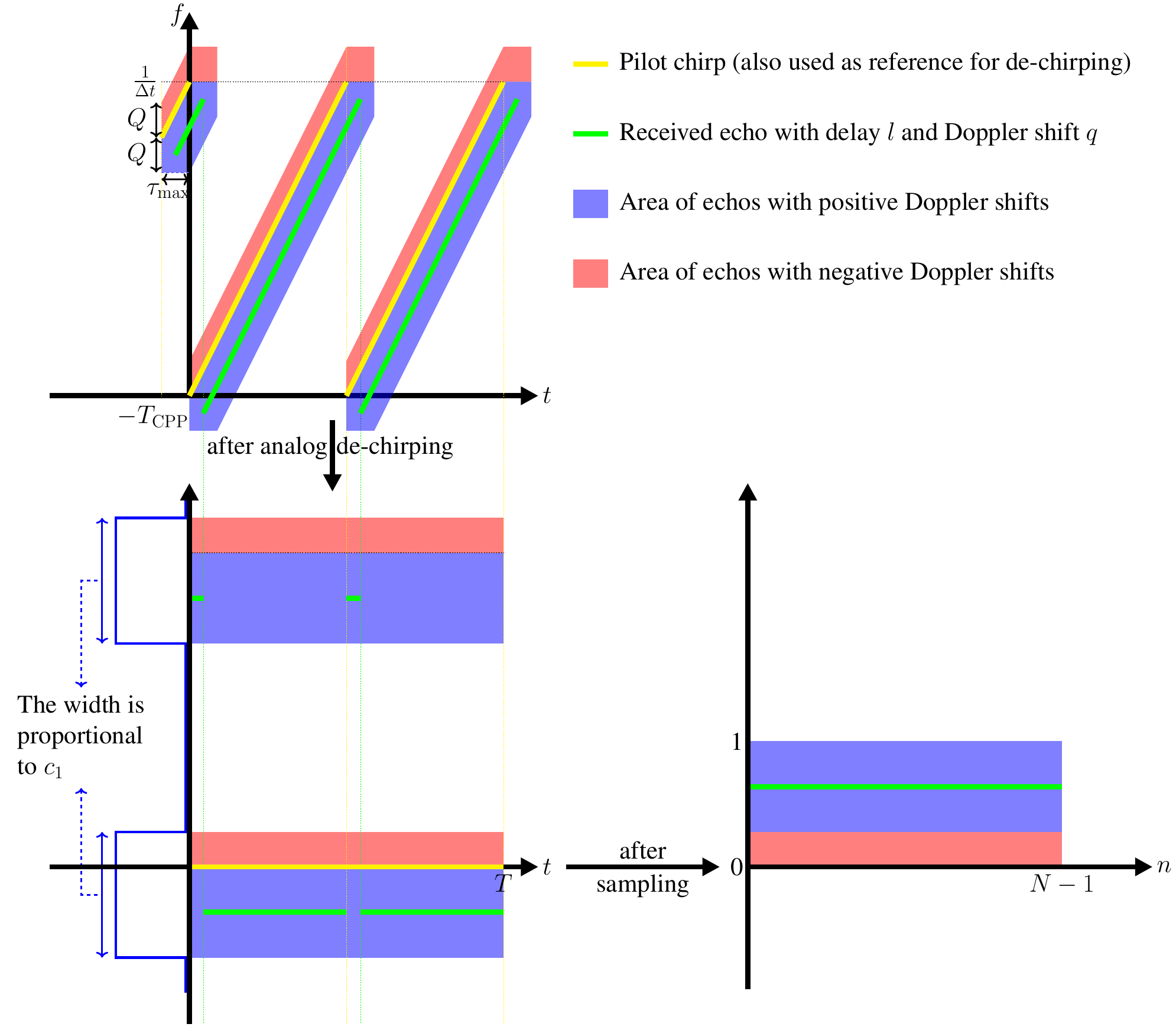}
    \caption{Time–frequency view of AFDM-based monostatic sensing with analog dechirping}
    \label{fig:monostatic_dechirping}
\end{figure}

However, one of the critical challenges in this setup, as previously mentioned, is managing SI, the direct leakage from the transmitter to the receiver caused by antenna coupling and imperfect isolation, as shown in Fig.~\ref{fig:SysModel}. This issue can be mitigated through analog dechirping, which suppresses the strong self-interference component before it reaches the ADC. To illustrate how analog dechirping operates, Fig.~\ref{fig:monostatic_dechirping} shows the time–frequency representations of the pilot chirp (also used as the reference for dechirping) in yellow, the received echo with delay $l$ and Doppler shift $q$ in green, the region corresponding to echoes with positive Doppler shifts in blue, and the region corresponding to echoes with negative Doppler shifts in red.

Due to the very short propagation delay associated with transmitter–receiver leakage, the strong SI signal from the pilot chirp coincides in the time–frequency domain with the pilot chirp itself. This SI signal must be suppressed before ADC sampling. By applying analog dechirping using a copy of the transmitted pilot as the reference signal, the SI collapses into a zero-frequency component, effectively becoming a direct-current (DC) signal. This component appears in yellow in Fig.~\ref{fig:monostatic_dechirping} and can be effectively removed using a simple \emph{DC-blocking module}. The desired target echoes, shown in green, retain their delay–Doppler information. The red region (whose length depends on the maximum Doppler shift) and the blue region (whose length depends on the maximum delay and the $c_1$ parameter) correspond to echoes with negative and positive Doppler shifts, respectively, spanning all resolvable delays up to the maximum delay. Due to the AFDM waveform’s structure, this spectral support is \emph{composite}, consisting of two disjoint frequency intervals created by discontinuities in the instantaneous frequency of the chirped signal. Capturing both intervals requires only a simple analog filter (highlighted in blue in Fig.~\ref{fig:monostatic_dechirping}). This antialiasing filter also removes the nonpilot part of the SI signal attributable to the data-carrying chirps associated with DAFT-domain indices reserved for data (see Fig.~\ref{fig:FrameSignlePilotWithData}). In short, SIC in AFDM-based monostatic sensing boils down to two lightweight steps: DC blocking and basic analog filtering.

Note that after analog-to-digital conversion at a sampling rate that is an integer fraction of the full bandwidth (i.e., a rational fraction that divides the bandwidth) and greater than the combined bandwidth of the red and blue regions around zero, the dechirped signal occupies, in digital frequency, only the equivalent of one blue region and one red region. This occurs because the frequency discontinuities present before sampling disappear due to spectrum folding. As a result, the sampling rate can be significantly lower than the full signal bandwidth, since the useful part of the dechirped signal occupies only a small portion of it. This observation enables uniform sub-Nyquist sampling without compromising delay–Doppler resolution. In doing so, it directly addresses one of the critical challenges in monostatic ISAC systems: the high-bandwidth requirement for high-resolution sensing. By reducing the ADC sampling rate without sacrificing sensing performance, this approach significantly improves energy efficiency and scalability for practical implementation.

\section{AFDM-based Bistatic Sensing Systems}
In contrast to the monostatic configuration, the bistatic ISAC setup separates the transmitter and receiver spatially, enabling passive sensing at the receiver side, typically a UE, as shown in Fig.~\ref{fig:SysModel}. In this scenario, the UE does not transmit any signal but instead leverages the downlink ISAC waveform broadcast by the base station to extract range and velocity information from surrounding objects. This passive radar approach is particularly attractive for applications where sensing capabilities must be embedded in resource- and power-constrained devices without requiring dedicated transmit chains.

As in the monostatic case, the transmitted ISAC signal is based on the AFDM waveform. Its inherent chirp-like structure and time–frequency sparsity make it naturally suited for passive sensing. However, because UEs typically have limited hardware resources and operate under strict power and complexity constraints, introducing specialized analog components, such as mixers, delay lines, or analog correlators, is impractical. To overcome these challenges, the sensing operation at the UE must be performed using standard hardware with minimal modifications. This motivates the need for a purely digital solution that can operate efficiently at sub-Nyquist sampling rates without sacrificing sensing accuracy or requiring changes to the analog front end. In the proposed framework, sub-Nyquist sampling is applied only to the pilot signal, without an accompanying data component, as shown in Fig.~\ref{fig:FrameSignlePilotWithoutData}. This structure avoids pilot–data interference after sub-Nyquist sampling and ensures reliable sensing performance. Schemes for sub-Nyquist sampling of AFDM with mixed data–pilot frames can be conceived\footnote{Provided that the data part is destined to a UE different from the one performing the sensing} using carefully designed DAFT-domain interleaved multiplexing, but are omitted here for clarity of exposition.

\begin{figure}
\centering
\begin{subfigure}{.45\textwidth}
 \includegraphics[width =\linewidth]{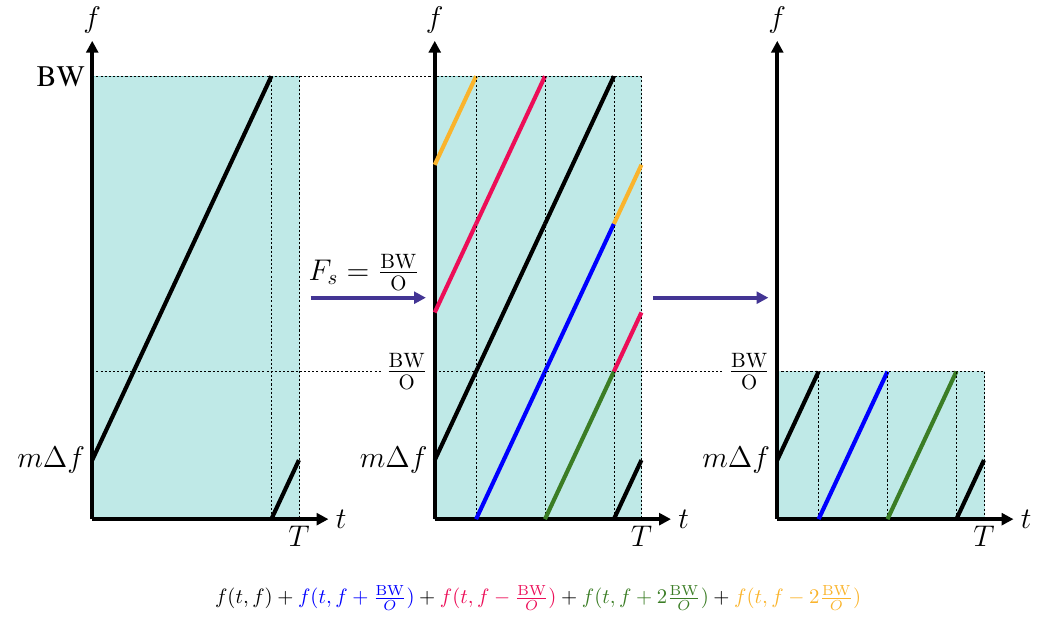}
    \caption{The effect of downsampling a chirp $\phi_m$ by a factor of $O=3$ on its time-frequency content}
\label{fig:ChirpPilot}
\end{subfigure}
\hspace{0.1\textwidth}
\begin{subfigure}{.45\textwidth}
    \centering
    \includegraphics[width=.5\linewidth]{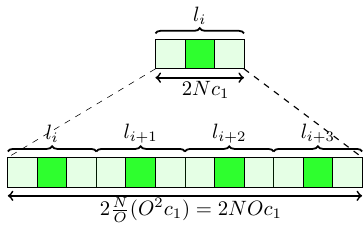}
\caption{The intuition behind why AFDM preserves delay resolution under sub-Nyquist sampling: $O = 4$ consecutive delay taps that would merge after downsampling become distinguishable as separate frequency offsets.}
    \label{fig:delayResDownSample}
\end{subfigure}
\caption{Sub-Nyquist sampling on AFDM signals} 
\end{figure}

To gain insight into the downsampling of a chirp, we first consider the $m$-th chirp $\phi_{m}[n]$. Let $O$ be a positive integer dividing $N$ (i.e., $O \mid N$). Downsampling $X[n]$ by a factor of $O$ yields the following $\tfrac{N}{O}$ samples:
\begin{equation}
\label{eq:phi_down}
\begin{aligned}
        & \phi_{\mathrm{down}, m}[k] \triangleq \phi_{m}[Ok] =  e^{\imath2\pi (c_1(Ok)^2 + c_2m^2 + \frac{mOk}{N})}\\
        &= e^{\imath2\pi\left((c_1O^2)k^2 + c_2m^2 + \frac{mk}{N/O}\right)}, \quad k = 0, \ldots, \frac{N}{O}-1.
\end{aligned}
\end{equation}
The downsampled chirp signal is thus a $\tfrac{N}{O}$-long linear chirp with chirp-rate parameters $(O^2 c_1, c_2)$ and a digital frequency shift of $m_{\tfrac{N}{O}}$, where the digital frequencies after downsampling cannot exceed $\tfrac{N}{O}-1$ due to spectrum folding, as shown in Fig.~\ref{fig:ChirpPilot}. This property suggests the possibility of using a $\tfrac{N}{O}$-point DAFT with parameters $(O^2 c_1, c_2)$ for sub-Nyquist AFDM receivers instead of the $N$-point DAFT with parameters $(c_1, c_2)$ used in the Nyquist-rate receiver. It can be proved that such a modified DAFT receiver, despite operating at a sampling rate $F_{\rm s} \triangleq \tfrac{\text{BW}}{O}$ and thus producing fewer samples, can still distinguish two delay taps separated by $\tfrac{1}{\text{BW}}$ seconds, even though this delay shift is smaller than $\tfrac{O}{\text{BW}}$ (the delay resolution after downsampling).

After transmitting the pilot signal $X[n]$ over a channel with maximum delay shift $l_{\max}$, the received samples are denoted by $r[n]$. If we downsample the received signal by a factor of $O$, the sequence $r[Ok]$ is, up to a phase rotation, equivalent to a linear combination of $\tfrac{N}{O}$-length linear chirps with chirp-rate parameters $(O^2 c_1, c_2)$, each shifted by a digital frequency offset $(m - 2N c_1){N/O}$. This observation motivates applying, at the receiver, a $\tfrac{N}{O}$-point DAFT with parameters $(O^2 c_1, c_2)$ to obtain the detection-domain samples $y{\mathrm{down}}[m]$.

It can be shown that all delay taps distinguishable using the Nyquist-rate DAFT outputs $y[m]$ are also distinguishable using the sub-Nyquist outputs $y_{\mathrm{down}}[m]$, where $y[m]$ are the $N$-point DAFT-domain samples of the received signal $r[n]$. However, the maximum unambiguous delay that can be estimated is reduced by the downsampling factor $O$ ($2Nc_1(l_{\max}+1) < \tfrac{N}{O}$). To understand why AFDM maintains delay resolution despite downsampling, it is useful to examine the parameters of the DAFT applied to the downsampled signal. Building on our previous work \cite{bemani2023affine_TWC, bemani2024integrated} on AFDM for linear time-varying (LTV) channels, it can be shown that a $\tfrac{N}{O}$-long AFDM signal based on a $(O^2 c_1,c_2)$ DAFT achieves the full diversity of an LTV channel with Doppler spread, in samples, equal to $2\tfrac{N}{O}O^2c_1 = 2NOc_1$. Consequently, every $O$ consecutive path delays that would otherwise merge into one peak due to $O$-fold downsampling are now distinguishable in sub-Nyquist AFDM, as they can be interpreted as $O$ different (and thus distinguishable) Doppler frequency shifts of the same delay tap, as illustrated in Fig.~\ref{fig:delayResDownSample}. AFDM therefore exploits the fact that Doppler (i.e., frequency) resolution is unaffected by downsampling to compensate for the loss in delay resolution.

Fig.~\ref{fig:TwoTargets}, with parameters given in Table~\ref{tab:setup}, shows $|y_{\rm down}[n]|$ as a function of delay for a two-target scenario, comparing the proposed downsampled AFDM pilot with a narrowband New Radio (NR) positioning reference signal (PRS). PRS is an OFDM-based reference signal introduced in 5G NR for positioning. Although originally designed for wideband operation to achieve high positioning accuracy, here it is restricted to the reduced bandwidth corresponding to the sub-Nyquist sampling rate. When the delay separation exceeds $O$ times the delay resolution ($O \cdot \tfrac{1}{\text{BW}} = 0.04~\mu$s), both the downsampled AFDM and the narrowband PRS can resolve the two targets. However, when the separation is smaller, specifically larger than the Nyquist delay resolution $\tfrac{1}{\text{BW}} = 0.01~\mu$s but smaller than $0.04~\mu$s, the proposed downsampled AFDM can still distinguish them, whereas the narrowband PRS merges them into a single detection.
\begin{figure}
\centering
\begin{subfigure}{.47\textwidth}
   \centering
    \includegraphics[width=\linewidth]{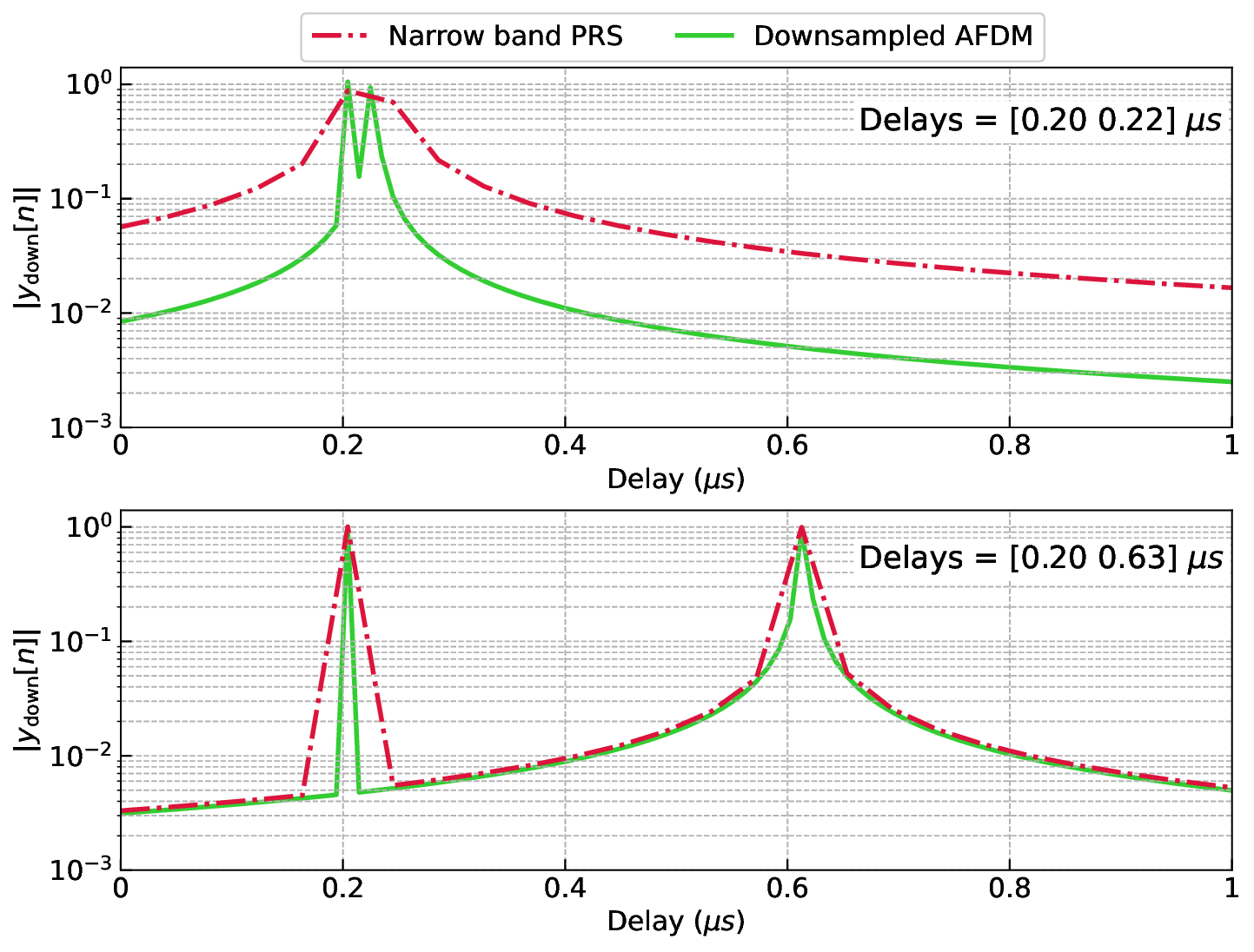}
    \caption{Detection output for two targets}
    \label{fig:TwoTargets}
    \label{fig:TwoTargets}
\end{subfigure}
\hspace{0.1\textwidth}
\begin{subfigure}{0.47\textwidth}
 \centering
    \includegraphics[width=\linewidth]{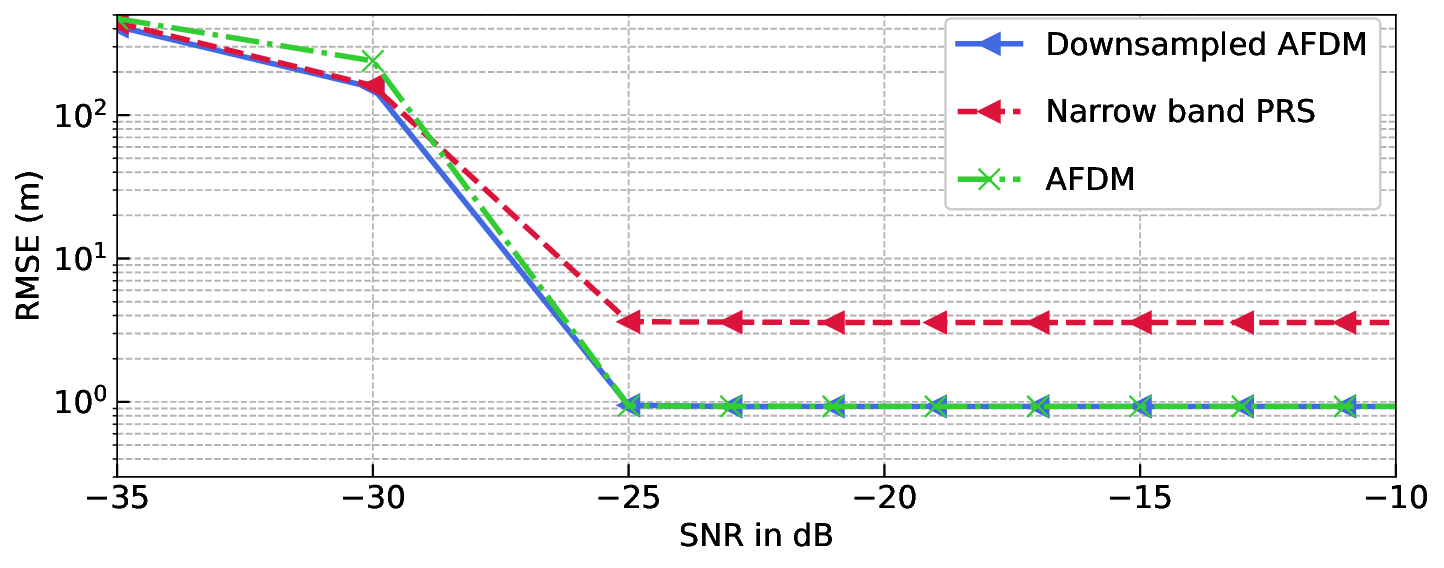}
    \caption{Range estimation performance of AFDM, downsampled AFDM and narrow-band PRS}
    \label{fig:RMSE}
\end{subfigure}

\caption{} 
\end{figure}

The loss of resolution for the narrowband PRS is further quantified in Fig.~\ref{fig:RMSE}, which compares the root-mean-square error (RMSE) of range estimation. Both AFDM and its downsampled version achieve the same RMSE saturation level, confirming that downsampling does not degrade resolution. In contrast, the narrowband PRS exhibits a higher saturation level, limiting its accuracy due to resolution loss. Specifically, its RMSE saturates at $\tfrac{1}{\sqrt{12}} \cdot \tfrac{O}{\text{BW}}$, whereas for the downsampled AFDM it remains at $\tfrac{1}{\sqrt{12}} \cdot \tfrac{1}{\text{BW}}$. This result highlights the ability of the downsampled AFDM to preserve range resolution.

\begin{table}
\centering

\caption{Simulations setting}
\begin{tabular}[t]{ll}
 \hline
 \textbf{Parameter} & \textbf{Value} \\ [0.5ex] 
 \hline\hline
$\#$ PRB for full-band PRS & 272 \\ \hline
AFDM frame size,$N$ & $\#$ PRB*12 = 3264\\ \hline
Subcarrier spacing, $\Delta f$ & 30KHz \\ \hline
FFT size & 4096 \\ \hline
Sequence type& Gold sequence \\ \hline
Parameters $(c_1, c_2)$& $(\frac{-1}{2N}, \frac{-1}{2N})$ \\ \hline
Bandwidth, $N\Delta f$ & 97.92 MHz \\ \hline
Delay resolution, $\frac{1}{BW}$ & 0.01 $\mu s$\\ \hline
Downsampling factor, $O$ & 4 \\ \hline
Number of trials & 100000 \\ \hline 
\end{tabular}
\label{tab:setup}
\end{table}%



\section{Design of Multichirp Pilot Signals for Sub-Nyquist Receivers}
As discussed earlier, the parameter $c_1$ should be kept small to avoid limiting the maximum delay that can be estimated when using sub-Nyquist sampling. A large value of $c_1$ restricts the system’s ability to resolve long delay paths, since the maximum unambiguous delay decreases as $c_1$ increases. This limitation becomes particularly critical in multicell settings. In conventional chirp-based pilot designs, multicell interference is typically mitigated by assigning different chirp-rate values to the pilots used in different cells. For example, in fourth-generation (4G) and fifth-generation (5G) wireless systems, Zadoff–Chu (Z-C) sequences are generated in different cells using different \emph{root values}\footnote{The root value of $N$-long Z-C sequences is closely related to $2Nc_1$ in AFDM}~\cite{dahlman20205g}, exploiting the low cross-correlation between chirps with distinct rates.

However, in the context of sub-Nyquist sensing, varying the chirp rate across cells increases $|c_1|$ for some pilots, which can significantly reduce the maximum resolvable delay—an undesirable effect for high-resolution sensing. This issue arises in both \emph{monostatic} and \emph{bistatic} AFDM-based ISAC systems:
\begin{itemize}
\item In the monostatic case, large $|c_1|$ values (``large'' relative to $\tfrac{1}{2N}$) reduce the detectable range after analog dechirping and also increase the required sampling rate, since the post-dechirping bandwidth grows with $|c_1|$.
\item In the bistatic case, the same limitation appears when the UE performs purely digital sub-Nyquist processing, thereby constraining its sensing range.
\end{itemize}

To address this, we propose a unified pilot design applicable to both configurations. Instead of relying on chirp-rate diversity for multicell interference mitigation, we exploit the \emph{multi-chirp} nature of the AFDM pilot by carefully selecting the DAFT-domain pilot symbols $\{x_m\}$ from cell-specific sequences with favorable auto- and cross-correlation properties, while keeping the chirp rate fixed (or within a small range) across cells, as illustrated in Fig.~\ref{fig:FrameMultiplePilotWithData} and Fig.~\ref{fig:FrameMultiplePilotWithoutData} for monostatic and bistatic systems, respectively. For example, $\{x_m\}_{m=0}^{M-1}$ can be chosen as a pseudorandom sequence with different initialization seeds in neighboring cells, ensuring low cross-correlation without altering $c_1$. 
This approach offers several advantages: it preserves a fixed chirp rate across all cells, thereby maintaining the maximum unambiguous delay; it provides robust multicell interference mitigation without requiring any hardware modifications; and it is fully compatible with both analog-assisted monostatic receivers and purely digital bistatic receivers operating at sub-Nyquist rates. In the monostatic case, however, analog dechirping produces an area around zero frequency rather than a single DC component. Since this area is suppressed by SIC, part of the target echoes may be lost, resulting in a blind region whose extent is controlled by the parameter $M$. 


\section{Conclusion}
This article has demonstrated that the AFDM waveform is a strong enabler of ISAC applications, offering an effective trade-off between performance and implementation complexity. In monostatic sensing at the base station, this is achieved through AFDM's compatibility with low-complexity SIC schemes and reduced sampling rates via analog dechirping. In bistatic sensing at the UE, AFDM supports sub-Nyquist sampling without additional hardware modifications while preserving delay resolution. In both cases, these benefits are further enhanced by effective multicell interference mitigation enabled by the resource-assignment flexibility of the AFDM transform domain. These features collectively establish AFDM as a compelling candidate for future ISAC systems.

\bibliographystyle{IEEEtran}
\bibliography{IEEEabrv,Citations}
\end{document}